\documentclass[preprint, review, authoryear, 12pt, 3p]{elsarticle}

\usepackage[table,xcdraw]{xcolor}
\usepackage{todonotes}
\usepackage{comment}
\usepackage{lipsum}
\usepackage{amsmath}
\usepackage[hidelinks]{hyperref}
\usepackage{graphicx}
\usepackage{subfig}
\usepackage{multirow}
\usepackage{lipsum}
\usepackage{multirow}
\usepackage{dirtytalk}
\usepackage{natbib}
\usepackage{caption}
\usepackage{adjustbox}
\usepackage{tabularx}
\usepackage{floatrow} 
\usepackage[vlines]{tabularht}
\usepackage{pbox}

\usepackage{float}
\floatstyle{plaintop}
\restylefloat{table}
\usepackage{rotating}
\usepackage[linesnumbered,ruled,vlined]{algorithm2e}
\SetKwInput{KwInput}{Input}                
\SetKwInput{KwOutput}{Output}              

\def\doubleunderline#1{\underline{\underline{#1}}}

\newcommand{\concat}{\ensuremath{+\!\!\!\!+\,}}

\newcommand{\lav}[1]{\textcolor{black}{#1}}

\newsavebox\MBox

\biboptions{sort&compress, semicolon, round}

\journal{arXiv}

\begin{document}

\begin{frontmatter}

\title{CVR-Net: A deep convolutional neural network for coronavirus recognition from chest radiography images}

\author[label1,label5]{Md. Kamrul Hasan\corref{cor1}\fnref{label3}}

\address[label1]{Department of Electrical and Electronic Engineering (EEE)}

\cortext[cor1]{I am corresponding author}
\fntext[label4]{Department of EEE, KUET, Khulna-9203, Bangladesh.}

\ead{m.k.hasan@eee.kuet.ac.bd}

\address[label5]{Khulna University of Engineering \& Technology (KUET)}

\author[label1,label5]{Md. Ashraful Alam}
\ead{ashrafulalam16e@gmail.com}

\author[label1,label5]{Md. Toufick E Elahi}
\ead{toufick1469@gmail.com}

\author[label1,label5]{Shidhartho Roy}
\ead{swapno15roy@gmail.com}

\author[label1,label5]{Sifat Redwan Wahid}
\ead{Sifat.Redwan17@gmail.com}

\begin{abstract}

\subsection*{Background and Objective}
The novel Coronavirus Disease $2019$ (COVID-$19$) is a global pandemic disease spreading rapidly around the world. A robust and automatic early recognition of COVID-$19$, via auxiliary Computer-aided Diagnostic \lav{(CAD)} tools, is essential for disease cure and control. \lav{The Artificial Intelligence (AI) assisted CAD system can be significant tool for the chest radiography images such as Computed Tomography (CT) and X-rays. } 
However, designing such an automated tool is challenging as a massive number of manually annotated datasets are not publicly available yet, which is the core requirement of supervised learning systems.

\subsection*{Methods} 
In this article, we propose a robust CNN-based network, called CVR-Net (Coronavirus Recognition Network), for the automatic recognition of the coronavirus from CT or X-ray images.
The proposed end-to-end CVR-Net is a multi-scale-multi-encoder ensemble model, where we have aggregated the outputs from two different encoders and their different scales to obtain the final prediction probability. 
We train and test the proposed CVR-Net on three different datasets, where the images have collected from different open-source repositories. We compare our proposed CVR-Net with state-of-the-art methods, which are trained and tested on the same datasets.

\subsection*{Results} 
We split three datasets into five different tasks, where each task has a different number of classes, to evaluate the multi-tasking CVR-Net. Our model achieves an overall F1-score \& accuracy of $0.997$ \& $0.998$; $0.963$ \& $0.964$; $0.816$ \& $0.820$; $0.961$ \& $0.961$; and  $0.780$ \& $0.780$, respectively, for task-$1$ to task-$5$.

\subsection*{Conclusion}
As the CVR-Net provides promising results on the small datasets, it can be an auspicious CAD tool for the diagnosis of coronavirus to assist the clinical practitioners and radiologists.
Our source codes and model are publicly available (\url{https://github.com/kamruleee51/CVR-Net}) for the research community for further improvements. \\

\end{abstract}

\begin{keyword}
Coronavirus  disease \sep  Chest  computed  tomography and X-ray \sep Convolutional neural networks \sep Ensembling classifier \sep CAD tools. 
\end{keyword}

\end{frontmatter}

\section{Introduction}
\label{introduction}
\subsection{Problem Presentation and Motivation}
\begin{figure*}[!ht]
  \centering
  \subfloat{\includegraphics[width=16cm, height= 9cm]{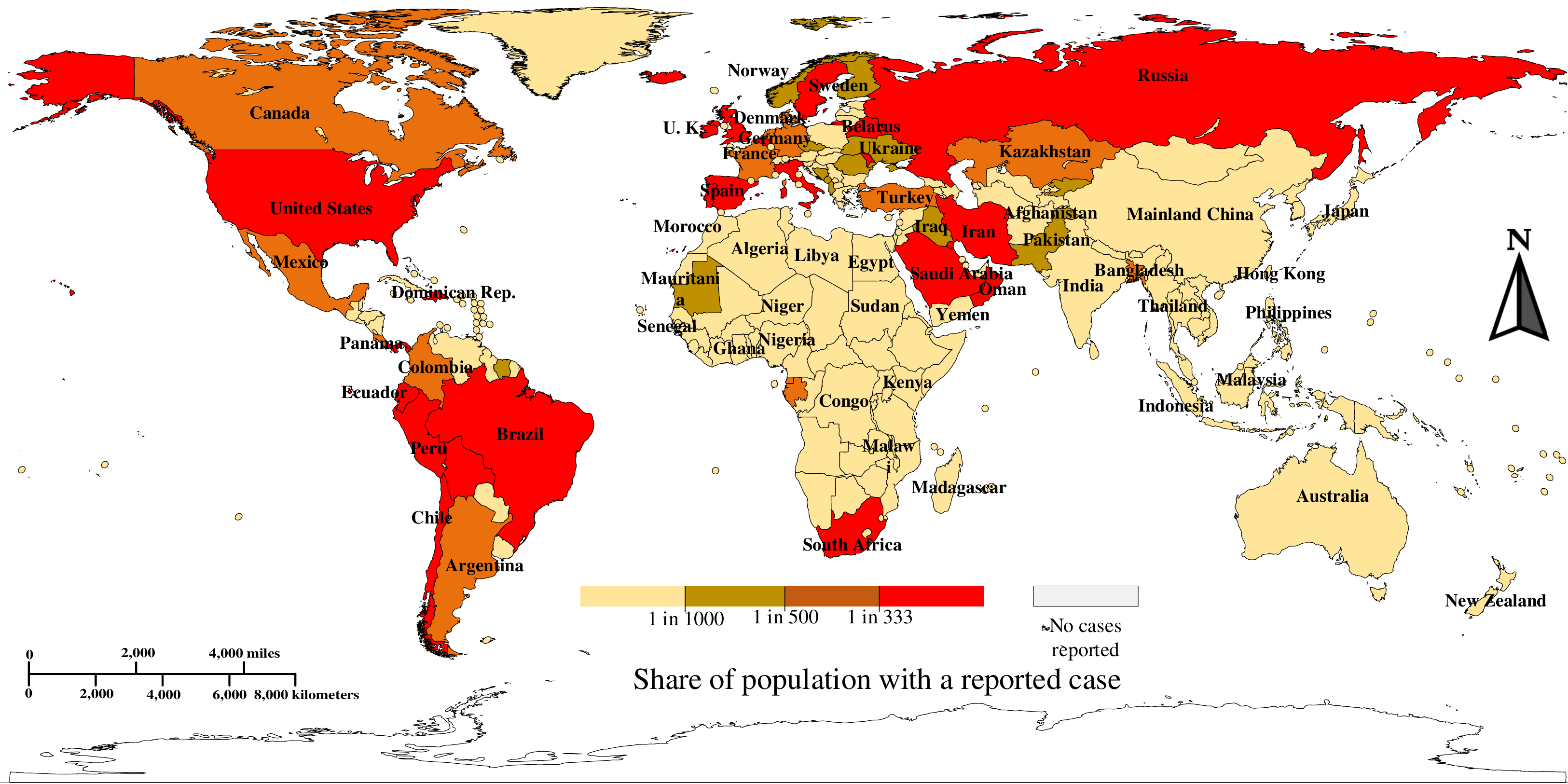}} 
  \caption{A world heat map of the corona pandemic per capita \citep{covid19global} [Accessed on $14$ July, $2020$].}
  \label{fig:Staistics}
\end{figure*}

\lav {A pneumonia of unknown cause detected in Wuhan, China was reported to the World Health Organization(WHO) office in China on 31 December, 2019 which was subsequently named  severe  acute  respiratory  syndrome  coronavirus  2  (SARS-CoV-2)  on  11  February,2020 as the virus causing the disease is genetically related to the corona virus responsible for the SARS outbreak of 2003.  The new disease was named as \say{COVID-19} by WHO on 11 February 2020  \citep{naming_covid}. As of July $2020$, the outbreak of $2019$ in Wuhan (China), has extended worldwide \citep{zhu2020novel, li2020early} with $12,286,264$ confirmed COVID-$19$ cases including $555,642$ deaths in a short period of 6 months (11 July 2020) \citep{who2020covid}, as presented in Fig.~\ref{fig:Staistics}}.
The clinical attributes, of severe COVID-$19$ epidemic, are bronchopneumonia that causes cough, fever, dyspnea, and subtle respiratory anxiety ailment \citep{wang2020clinical, chen2020epidemiological, li2020early}. A person with fever, cough, and influenza symptoms, is usually screened by the conventional methods of clinical trials, laboratory testing, and chest radiography to rule out pneumonia. Reverse Transcription Polymerase Chain Reaction (RT-PCR) is a commonly employed clinical screening test for COVID-$19$ infection, using respiratory specimens. Generally, radiologists apply the RT-PCR test as a base tool for detecting coronavirus, but it is a manual, complicated, tedious, and time-consuming procedure with a true-positive rate of $63.0\%$ \citep{wang2020detection}. There is a significant lack of inventory, leading to a delay in efforts to prevent and cure coronavirus disease \citep{yang2020point}. Many countries face crises with the incorrect number of positive COVID-$19$ cases due not only to the lack of test kits but also to the defer in the test results \citep{aj2020covid}. Such delays may cause affected patients to interact with and affect healthy people in the process. Furthermore, the RT-PCR kit is estimated to cost around $120\sim 130$ USD, and requires a specially designed biosafety laboratory to house the PCR unit, each of which can cost $15,000\sim 90,000$ USD  \citep{aj2020corona}. Nevertheless, using a costly screening device with delayed test results allows it to spread and worsen the situation. This scenario is a problem for low-income countries, but some developed countries also struggle to alleviate that limitation \citep{wetsman2020coronavirus}.


It is observed that most of the COVID-$19$ cases have common characteristics on radiographic images, such as CT and X-ray, including bilateral, multi-focal, ground-glass opacities with a peripheral or posterior distribution, mainly in the lower lobes, early- and late-stage pulmonary consolidation \citep{huang2020clinical, corman2020detection,xu2020deep,singh2020classification}. Those images can be utilized to develop a sensitive CAD tool to detect COVID-$19$ pneumonia and be considered a screening tool with RT-PRC \citep{lee2020covid}.
Application of machine learning methods for automatic diagnosis in
the medical field, via CAD tools, has recently gained popularity by becoming an adjunct tool for the clinicians \citep{litjens2017survey,ker2017deep,shen2017deep}. The deep CNN-based published methods, for automatic coronavirus recognition, are briefly presented in the next subsection. 



\subsection{Recent Methods}
\citet{ozturk2020automated} proposed a model, called DarkCovidNet having $17$ convolutional layers, for binary classification (COVID vs No-Findings) and multi-class classification (COVID vs No-Findings vs Pneumonia) using chest X-ray images. The author employed the DarkNet \citep{redmon2017yolo9000} model for the You Only Look Once (YOLO) \citep{redmon2017yolo9000} real-time coronavirus detection system. 
A Deep CNN model, called CoroNet, automatically detect COVID-$19$ infection from chest X-ray images, which was proposed by \citet{khan2020coronet}. 
The CoroNet was based on pre-trained Xception architecture \citep{chollet2017xception} in ImageNet \citep{imagenet}, with a dropout layer \citep{srivastava2014dropout} and two fully-connected layers. They trained and evaluated their model on different tasks like binary and multi-class recognition. 
\citet{ghoshal2020estimating} investigated uncertainty of the coronavirus classification report, using the drop-weights-based Bayesian CNN, as the availability of uncertainty-aware Deep Learning (DL) can ensure more extensive adoption of DL in clinical applications.
\cite{narin2020automatic} implemented three different deep CNN models such as ResNet-$50$ \citep{he2016deep}, InceptionV3 \citep{szegedy2017inception}, and Inception-ResNetV$2$ \citep{szegedy2017inception}, where they also used transfer learning for the detection of coronavirus and pneumonia infected patient. The authors showed that chest X-ray images and ResNet-$50$ are the best tools for the detection of COVID-$19$. 
\cite{hemdan2020covidx} proposed a DL framework, called COVIDX-Net, where they experimented on seven different CNN architectures, such as VGG-$19$ \citep{simonyan2014very}, DenseNet-$121$ \citep{huang2017densely}, ResNetV$2$ \citep{he2016identity}, InceptionV$3$, Inception-ResNetV$2$, Xception, and MobileNetV$2$ \citep{howard2017mobilenets}. In the end, their results suggested that VGG-$19$ and DenseNet-$201$ are better for coronavirus screening systems with X-ray images.
Three comparatively shallow networks, such as MobileNetV$2$, SqueezeNet \citep{lecun2010convolutional}, and ResNet-$18$, and five deep networks, such as InceptionV$3$, ResNet-$101$, CheXNet \citep{rajpurkar2018deep}, VGG-$19$, and DenseNet-$201$, were trained and evaluated by \cite{chowdhury2020can} for coronavirus recognition. Their study experimentally showed that DenseNet-$201$ outperformed other deep CNN networks, while the authors trained their model with different image augmentations.
\cite{abbas2020classification} proposed a framework by adopting a deep CNN, called Decompose, Transfer, and Compose (DeTraC) \citep{abbas2020detrac} for the classification of COVID-$19$ chest X-ray images, where the authors implemented the DeTraC in two phases. Firstly, they trained, using a gradient descent optimization, the backbone pre-trained CNN model of DeTraC to extract deep local features from each image. Secondly, they used the class-composition layer of DeTraC to refine the final classification of the images. 
\cite{zhao2020covid} developed diagnosis methods based on multi-task learning and self-supervised learning, where the authors proposed an open-source COVID dataset of CT images with a binary class (COVID and NON-COVID). For the classification task, they train DenseNet-$169$ and ResNet-$50$, via a pre-trained model on ImageNet weights, with their newly proposed dataset. 
\citet{hall2020finding} explored the usefulness of the chest X-ray
images with various deep CNN models for the diagnosis of the COVID-$19$ disease. The authors employed pre-trained ResNet-$50$ and VGG-$16$ \citep{simonyan2014very} on ImageNet. 
\citet{afshar2020covid} proposed a CNN model named COVID-CAPS, which was based on the Capsule Networks (CapsNets) for handling the small datasets of coronavirus. CapsNets are alternative models of CNN, which are capable of capturing spatial information using routing by agreement, through which capsules try to reach a mutual agreement on the existence of the objects.
Their proposed COVID-CAPS model had $4$ convolutional layers and $3$ capsule layers, where batch normalization \citep{ioffe2015batch} followed the former layers. The authors fine-tuned all the capsule layers, while the conventional layers were frozen with pre-trained weights of ImageNet.
\citet{apostolopoulos2020covid} accomplished comprehensive experiments on state-of-the-art CNN models applying transfer learning. In the end, the authors found that VGG-$19$ outperforms other CNNs for accuracy, while MobileNetv$2$ outperforms VGG-$19$ in terms of specificity.
\citet{he2020benchmarking} built a COVID CT dataset, called China Consortium of Chest CT Image Investigation (CC-CCII), with three classes: novel coronavirus pneumonia, common pneumonia, and healthy controls. The authors trained $3$D DenseNet3D-$121$ on their proposed CC-CCII dataset, and they experimentally validated that $3$D CNNs outperform $2$D CNNs in general. 
\citet{singh2020classification} implemented a CNN-based model named multi-objective differential evolution–based CNN for the classification of COVID-$19$. They fine-tuned the parameters of the CNN model using a multi-objective fitness function. The differential evolution algorithm was used for the optimization of the multi-objective fitness function. In differential evolution, the model was optimized iteratively using mutation, crossover, and selection operation to determine the best available solution.
\citet{apostolopoulos2020extracting} extracted massive high-dimensional features, using a pre-trained MobileNetV$2$ architecture, corresponding to six diseases. Finally, they used fully-connected layers to classify those features for the identification of the coronavirus. 
\citet{farooq2020covid} employed ResNet-$50$ using transfer learning with progressively resizing the input images to $128\times 128 \times 3$, $224 \times 224 \times 3$, and $229 \times 229 \times 3$ pixels, where the authors also fine-tuned the network at each stage.
\citet{ozkaya2020coronavirus} extracted deep features using VGG-$16$, GoogleNet \citep{szegedy2015going}, and ResNet-$50$ models, which were classified by Support Vector Machine (SVM) \citep{furey2000support} with linear kernel function. They also applied the t-test method to reduce the feature dimension for reducing the overfitting. 
\citet{rajaraman2020iteratively} evaluated ImageNet pre-trained CNN models such as VGG-$16$,  VGG-$19$, InceptionV$3$, Xception, Inception-ResNetV$2$, MobileNetV$2$, DenseNet-$201$, and NasNet-mobile \citep{pham2018efficient}. 
Then, they optimized the hyperparameters of the CNNs using a randomized grid search method \citep{bergstra2012random}. In the end, the authors proposed an ensemble of those CNN models for the final coronavirus recognition. 
\citet{tougaccar2020covid} restructured the data classes using a fuzzy color technique, where they stacked a structured image with the original images. The authors trained MobileNetV$2$ and SqueezeNet to extract the deep features, which were then processed using the social mimic optimization method \citep{balochian2019social}. After that, selected features were combined and classified using the SVM for the recognition of coronavirus.


\subsection{Our Contribution}
The discussions mentioned above, on automated coronavirus recognition systems, show that the deep CNN approaches are commonly employed methods than the different systems that rely on handcrafted features.
The former approaches provide auspicious reproducibility of results and amplify the speed of the diagnosis while being end-to-end methods.  
While many approaches have already been developed and implemented for coronavirus recognition, there is still room for performance improvement for different datasets.
However, it is very impractical to guesstimate the amount of depth of the CNN networks and the times of subsampling, when utilizing datasets are small in size.  
In this article, we propose an end-to-end coronavirus recognition network called CVR-Net, where we ensemble different scaled feature maps from different encoders through fully-connected layers. Such an ensembling allows the network to access different depths and scales of the feature maps of different encoders for generating the final prediction. In our CVR-Net, the newly added depth and subsampling can not degrade the final prediction as their previous depths and subsampling compensate them.
To overcome the overfitting and build a generic CVR-Net, with the limited datasets, we apply geometry-based image augmentations and transfer learning on ImageNet \citep{krizhevsky2012imagenet}. 
Besides, we also rebalance the imbalanced class distribution, as a massive number of positive coronavirus images are not available yet due to the recent COVID-$19$ pandemic.  
We validate our multi-tasking CVR-Net on three different datasets, of two different modalities, such as CT and X-ray, with a different number of classes (see in Table~\ref{tab:dataset}). We have collected images from different open-sources, such as Kaggle, GitHub, and grand challenges (see in subsection \ref{Datasets}).   
To our best knowledge, the proposed CVR-Net has achieved state-of-the-art results on three different datasets, having a different number of classes, while being an end-to-end coronavirus diagnosis system.


The rest of the paper is structured accordingly. We explain the proposed framework for the recognition of coronavirus and datasets in section \ref{materialandMethods}. The results and different experiments are reported in section \ref{ResultsandDiscussion}. We interpret the obtained results from the proposed CVR-Net in section \ref{Discussion}. Finally, we conclude this paper in section \ref{Conclusion}.


\section{Materials and Methods}
\label{materialandMethods}
This section presents the materials and methods for conducting this research. Subsection \ref{Datasets} briefly describes utilized dataset. The designing of the proposed network (CVR-Net) is explained in subsection \ref{COVIDNet}. Finally, subsection \ref{Training} describes the training protocol of our network and the evaluation metrics.

\subsection{Dataset and Hardware}
\label{Datasets}
\begin{figure*}[!ht]
\centering
\subfloat[]{\label{a}\includegraphics[width=4cm,height=4.cm]{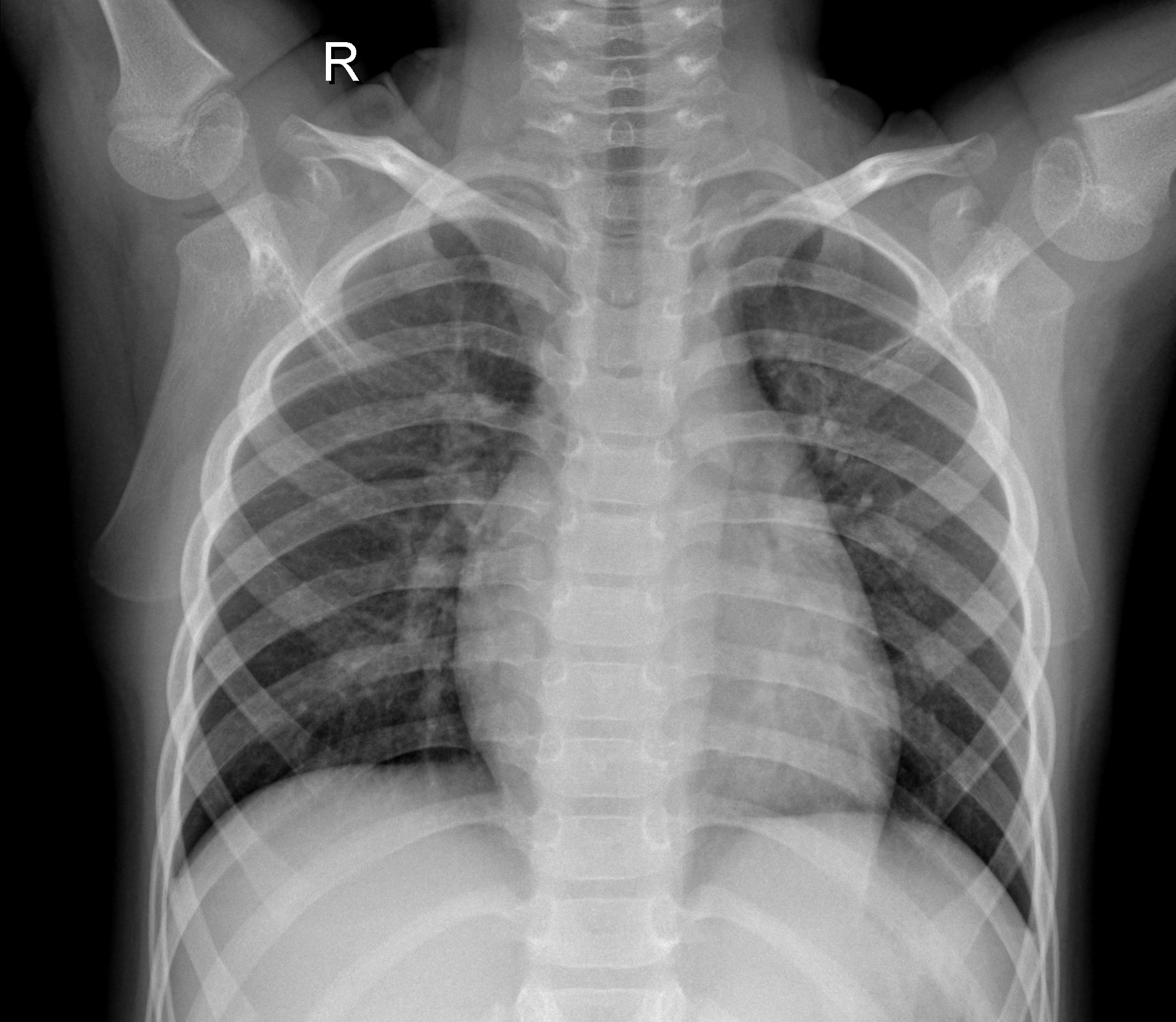}}
\hspace{0.2cm}
\subfloat[]{\label{a}\includegraphics[width=4cm,height=4cm]{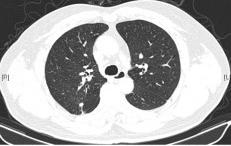}}
\hspace{0.2cm}
\subfloat[]{\label{a}\includegraphics[width=4cm,height=4cm]{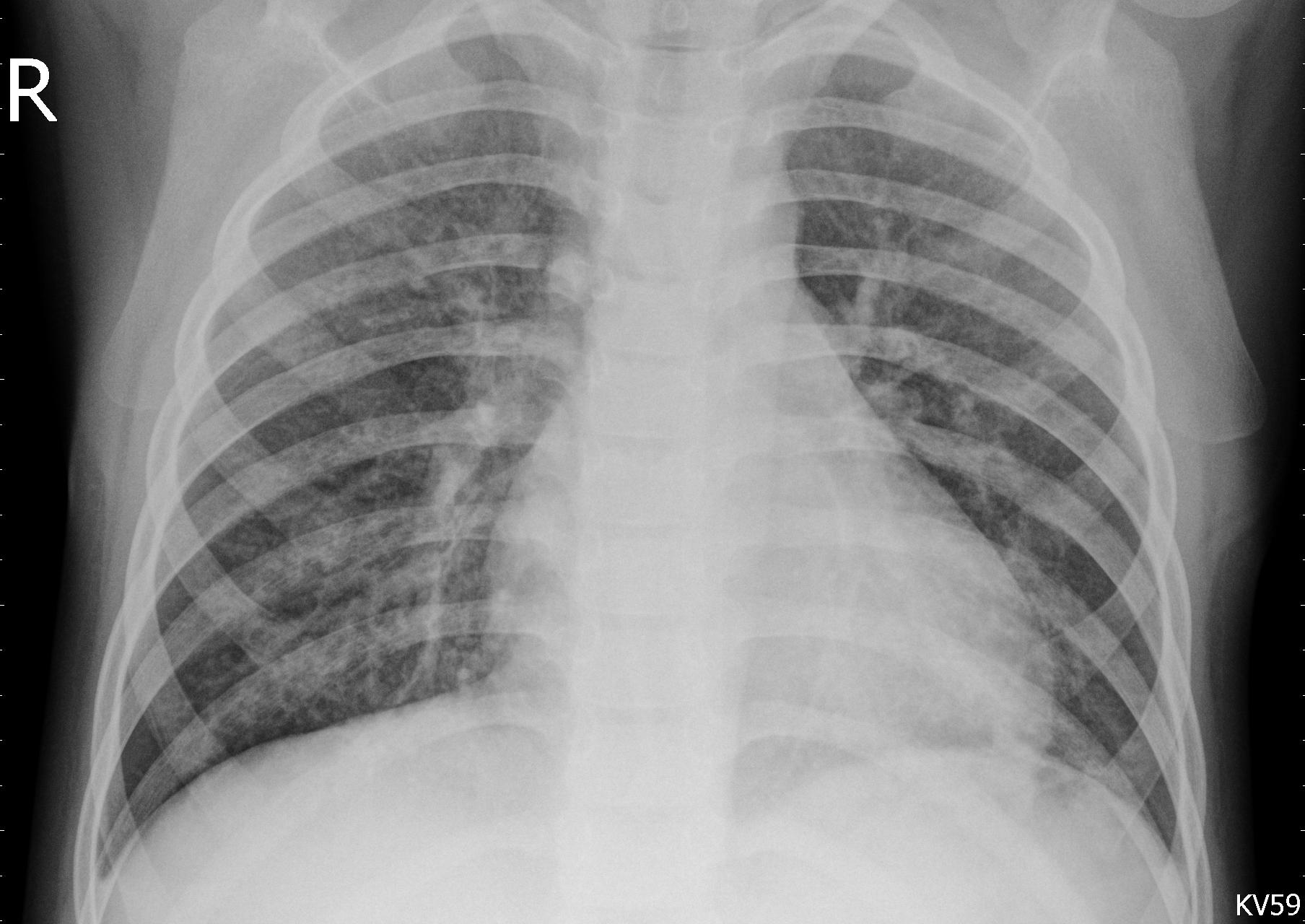}} \\
\subfloat[]{\label{a}\includegraphics[width=4cm,height=4cm]{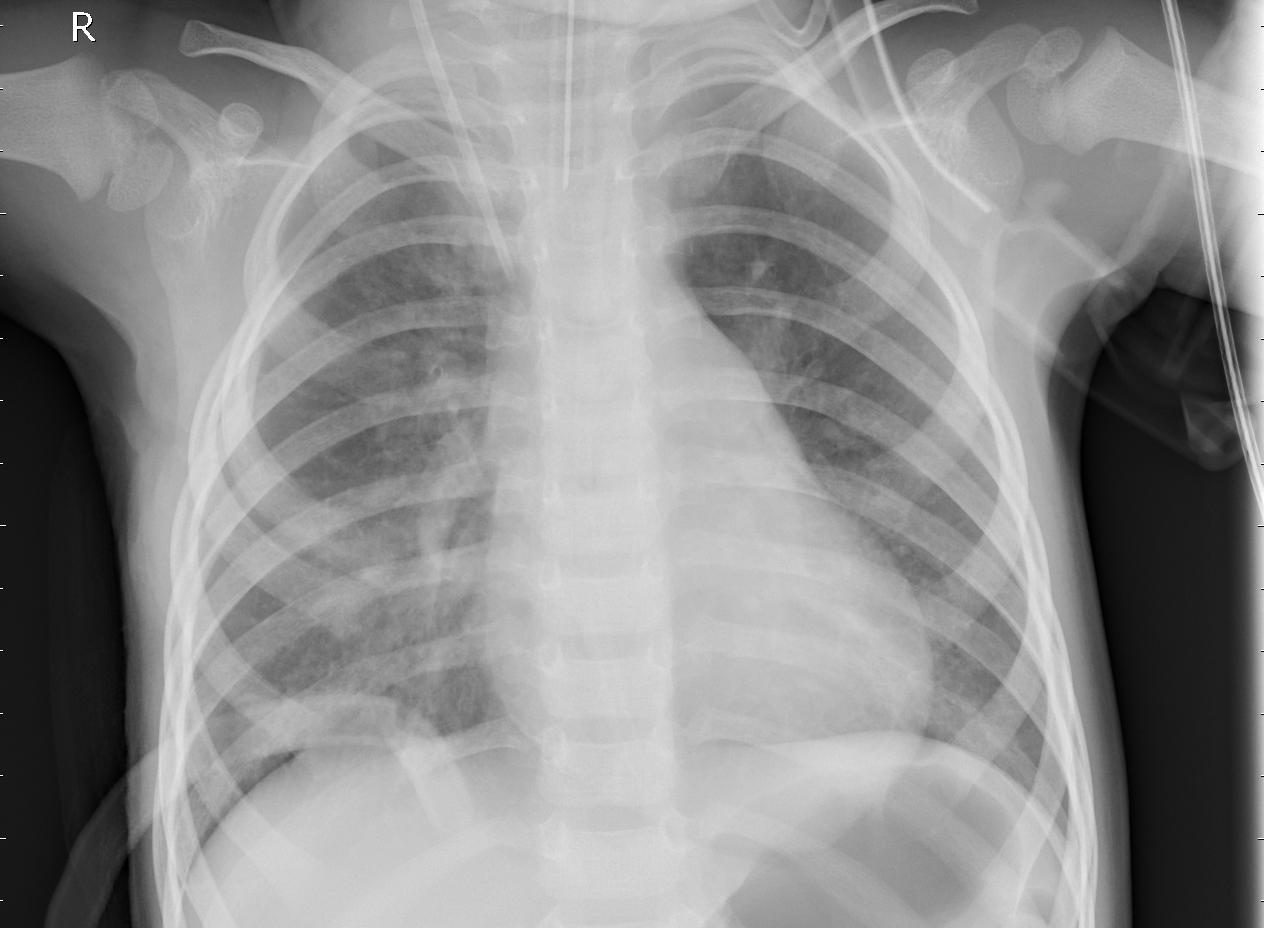}}
\hspace{0.2cm}
\subfloat[]{\label{a}\includegraphics[width=4cm,height=4cm]{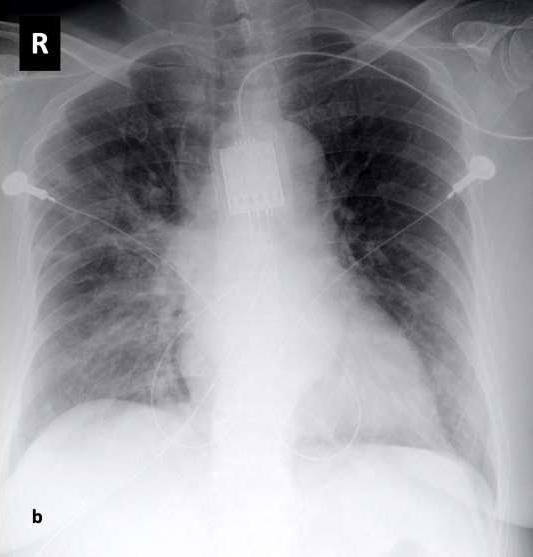}}
\hspace{0.2cm}
\subfloat[]{\label{a}\includegraphics[width=4cm,height=4cm]{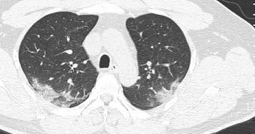}}
\caption{Samples of chest radiography images from the utilized datasets (a) Normal (X-ray), (b) Normal (CT), (c) Pneumonia viral, (d) Pneumonia bacterial, (e) COVID-19 (X-ray), and (f) COVID-19 (CT).}
\label{fig:sample}
\end{figure*}

\begin{figure*}[!ht]
\centering
  \includegraphics[width=12cm,keepaspectratio=True]{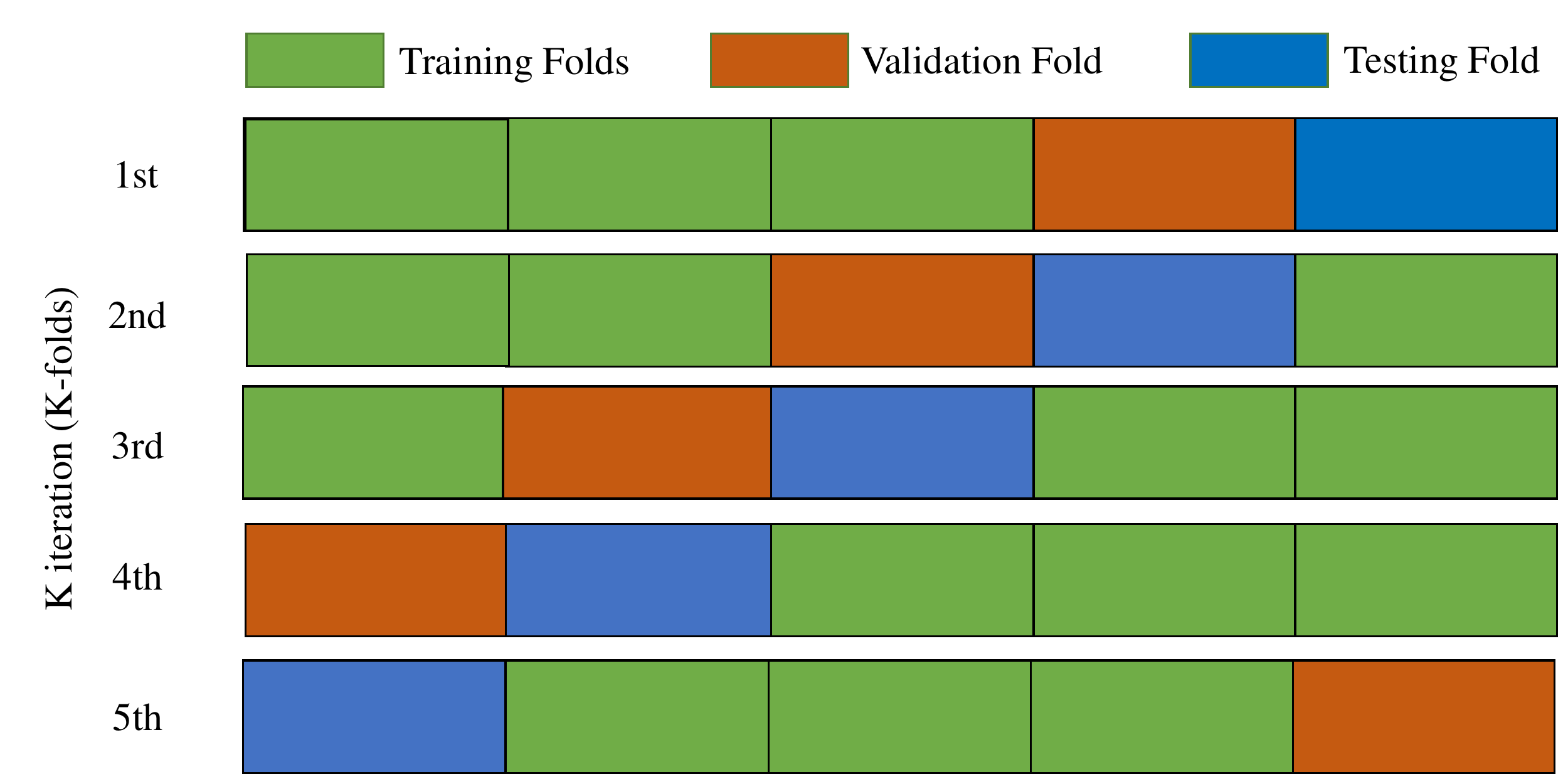}
  \caption{The partitioning of the datasets into five parts for selecting training, validation, and testing images.}
  \label{fig:kcv}
\end{figure*}

We train and evaluate our multi-tasking CVR-Net on three datasets, CT and X-ray images, with a different number of classes. We evaluate the proposed network on three different types of tasks, such as: healthy vs. coronavirus ($2$-class), healthy vs. pneumonia vs. coronavirus ($3$-class), and healthy vs. bacterial pneumonia vs. viral pneumonia vs. coronavirus ($4$-class). 
As COVID-19 is the recent pandemic all over the world, there is still a lack of suitable annotated public datasets as it requires experts for labeling. However, we collected the positive COVID-19 X-ray images from an open-source GitHub\footnote{https://github.com/ieee8023/covid-chestxray-dataset} repository of \citet{cohen2020covid}, the authors compiled the images from various authentic sources (Radiological Society of North America (RSNA), Radiopaedia, \textit{etc}). 
We collect the Pneumonia (both bacterial \& viral) and healthy chest X-ray images from Kaggle repository \say{Chest X-Ray Images (Pneumonia)} \citep{ChestXRa44}. 
These two datasets are merged for dataset-$1$, as utilized in CoroNet by \citet{khan2020coronet}, which has three different tasks, as presented in Table~\ref{tab:dataset}. 
The utilized dataset-$2$ is the combination of dataset-$1$ and additional images from another Kaggle repository \say{Pneumonia sample X-Rays} \citep{Pneumoni29}, as it was utilized by \citet{tougaccar2020covid}.
Finally, the dataset-$3$ is collected from COVID-$19$ grand challenges \citep{zhao2020covid}, which is the CT images of COVID and healthy patients and collected from Tongji Hospital, Wuhan, China. 
The distribution of all the three datasets is presented in Table~\ref{tab:dataset}, where we assign three different tasks, with a different number of classes, for dataset-$1$ and single task for the other two datasets.   
\begin{table*}[!ht]
\caption{Utilized data distribution of five different tasks, of three datasets, to validate our proposed CVR-Net, where data was accumulated from different open-sources.}
\centering
\footnotesize
\begin{tabular}{llll}
\hline
\rowcolor[HTML]{C0C0C0} 
Datasets                    & Task Types                        & Class categories & No. of Images \\ \hline
                            &                                   &  Normal (NOR)                 &    $5,856$            \\
                            & \multirow{-2}{*}{Task-1: 2-class} &  Novel Corona Positive (NCP)   & $500$               \\ \cline{2-4} 
                            &                                   & Normal (NOR)      &   $1,583$           \\
                            &                                   & Common Pneumonia (CPN)        &  $4,273$             \\
                            & \multirow{-3}{*}{Task-2: 3-class} & Novel Corona Positive (NCP)    &  $500$              \\ \cline{2-4} 
                            &                                   & Normal (NOR)                 &    $1,583$           \\
                            &                                   & Common Pneumonia Bacterial (CPB)     &         $2780$      \\
                            &                                   & Common Pneumonia Viral (CPV)         &            $1493$   \\
\multirow{-9}{*}{Dataset-1} & \multirow{-4}{*}{Task-3: 4-class} & Novel Corona Positive (NCP)          &    $500$           \\ \hline
                            &                                 &       Normal (NOR)             &    $1,648$           \\
                            &                                   &   Common Pneumonia (CPN)                 &    $4,371$           \\
\multirow{-3}{*}{Dataset-2} & \multirow{-3}{*}{Task-4: 3-class} &          Novel Corona Positive (NCP)        & $500$              \\ \hline
                            &                                   &     Normal (NOR)           &    Train/test = $292/105$           \\
\multirow{-2}{*}{Dataset-3} & \multirow{-2}{*}{Task-5: 2-class} &     Novel Corona Positive (NCP)               &   Train/test = $251/98$            \\ \hline
\end{tabular}
\label{tab:dataset}
\end{table*}
Several example of CT and X-ray images for different classes is presented in Fig.~\ref{fig:sample}. 
We have applied a $5$-folds cross-validation technique, as presented in Fig.~\ref{fig:kcv}, for the first two datasets to select training, validation, and testing images. 
The class-distribution of all the datasets, as shown in Table~\ref{tab:dataset}, demonstrates that the images are imbalanced, which makes the classifier to be biased to the particular class having more samples. However, we have employed class rebalancing techniques by penalizing the majority class's loss to build a generic classifier even though datasets are imbalanced.

The models were implemented using the Python programming language with different Python and Keras APIs \citep{chollet2015keras} and the experiments were carried out on a machine running \textit{Windows-10} operating system with the following hardware configuration:
Intel\textsuperscript{\tiny\textregistered} Core\textsuperscript{\tiny{TM}} i$7$-$7700$ HQ CPU @ $3.60\,GHz$ processor with Install memory (RAM): $32.0\,GB$ and GeForce GTX $1080$ GPU with $8\,GB$ 
memory.


\subsection{Proposed CVR-Net Architecture}
\label{COVIDNet}
Efficaciously classifications, of medical images, have an essential role in aiding clinical cure and control. For instance, an X-ray investigation for diagnosing pneumonia is the best approach \citep{world2001standardization}, but it needs professional radiologists or experts, which is a rare, costly, and arduous for some regions.
The employment of the conventional machine learning algorithms, in medical image classification, began long ago \citep{yadav2019deep}, which has several disadvantages, such as the poor performance than the practical standard; the implementation of them is quite slow; the extraction and selection of the features are time-consuming and tedious; fluctuate a lot according to various objects \citep{kermany2018identifying}. 
The deep neural networks, notably the CNNs, are comprehensively applied for image classification or recognition, which have earned powerful performance since $2012$ \citep{rawat2017deep}. Recently, CNN-based medical image classification rivals human expertise. For instance, CheXNet, a CNN classifier trained on a chest X-rays dataset with more than $100,000$ frontal-views, achieved better results than the average performance of four experts. 
Moreover, \citet{kermany2018identifying} proposed a CNN-based classifier, with a transfer learning, to recognize $108,309$ optical coherence tomography images, where the average error, from the CNN model, was equivalent the errors from $6$ different human experts.
Currently, the automatic recognition of coronavirus is one of the critical topics for the researcher, where images are hard to accumulate, as the collection and annotation of COVID-$19$ data are time-consuming, costly, and required expert explanations.

\begin{figure*}[!ht]
  \centering
  \subfloat{\includegraphics[width=16cm, height= 8.5cm]{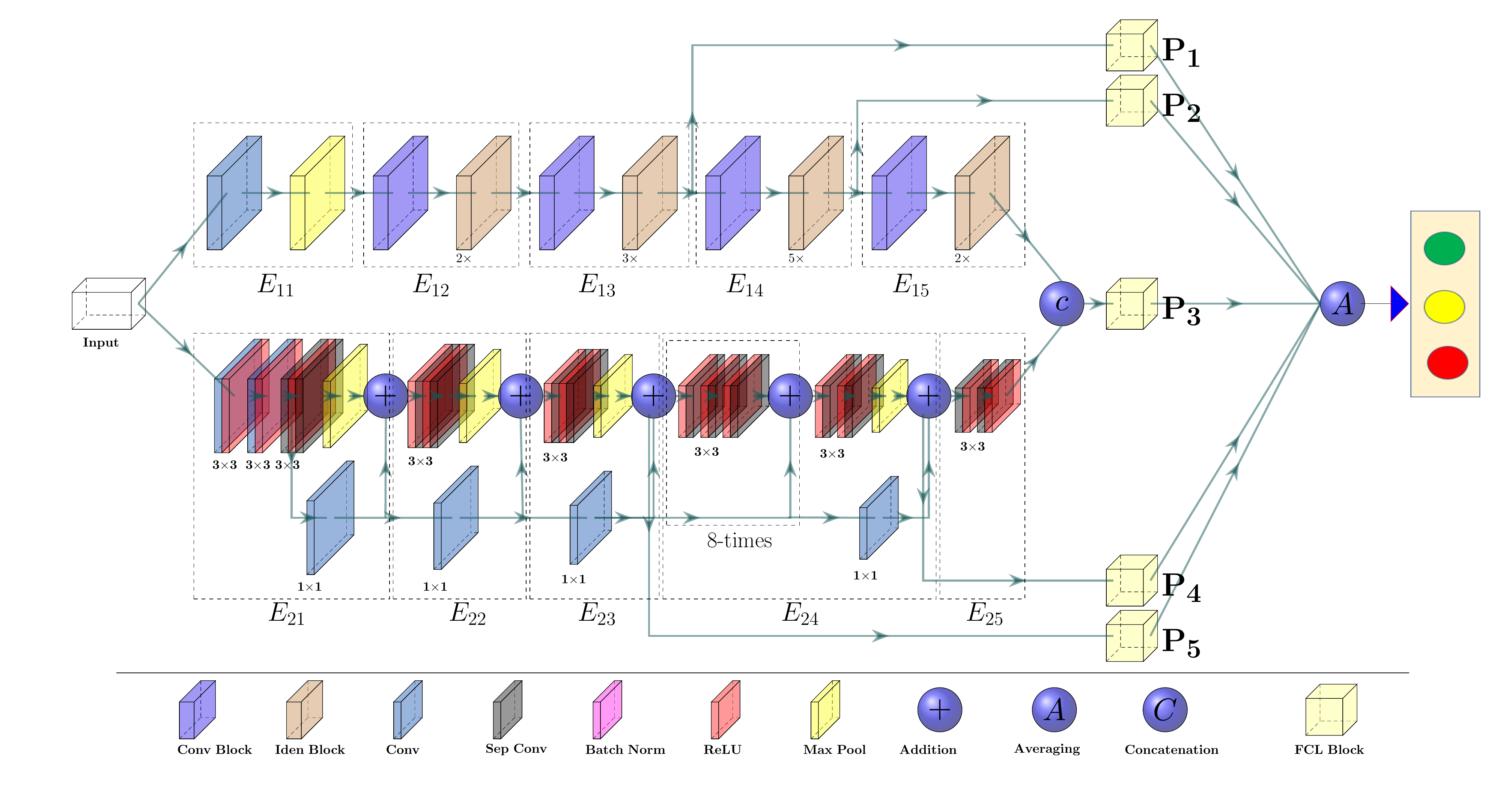}} 
  \caption{The proposed network, called CVR-Net, for the automatic coronavirus recognition from radiography images, where we ensemble the multi-encoder and multi-scale of the network, via fully connected blocks, obtain final recognition probability.}
  \label{fig:network}
\end{figure*}

\begin{figure*}[!ht]
  \centering
  \subfloat{\includegraphics[width=16cm, height= 4cm]{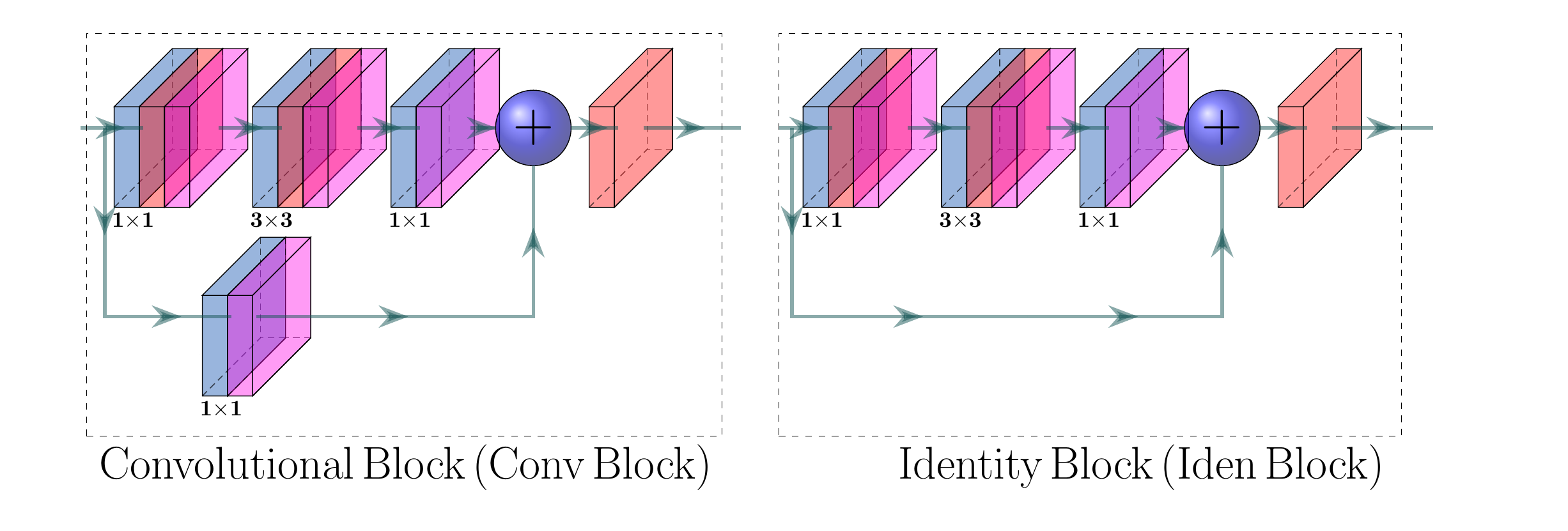}} 
  \caption{The convolutional (left) and residual (right) blocks \citep{he2016deep} of the proposed CVR-Net, where the output map is the summation of the input map and the generated map from the process (convolutions).}
  \label{fig:blocks}
\end{figure*}

However, designing an end-to-end recognition system is a challenging task as the CNNs may be indirectly limited when employed with highly variable and distinctive image datasets with limited samples such as COVID datasets. Moreover, individual CNN architecture may have different capabilities to characterize or represent the image data, which is often linked to a network's depth \citep{kumar2016ensemble}. The number of layers with increasing depth and amount of subsampling (a downsampling in pooling layers) is also challenging to guesstimate with the limited datasets. However, in this context, we propose a CNN-based end-to-end multi-tasking network, where we apply multi-encoder and multi-scale ensembling, as depicted in Fig.~\ref{fig:network}.
The proposed CVR-Net consists of two encoders, with the same input images, where each of the encoders has five blocks, namely $E_{1n}$ and $E_{2n}$, $n=1, 2, ..., 5$, for encoder-$1$ and encoder-$2$, respectively.  
The encoder-$1$ consists of the residual and convolutional blocks \citep{he2016deep}, as presented in Fig.~\ref{fig:blocks}, where the residual connections allow the information to flow or skip.
The residual connections, also known as skip connections, allow gradients to flow through a network directly, without passing through non-linear activation functions and thus avoiding the problem of vanishing gradients in the proposed CVR-Net \citep{he2016deep}. In residual connections, the output of a weight layer series is added to the original input and then passed through the non-linear activation function, as shown in Fig.~\ref{fig:blocks}.
However, in encoder-$1$, $7\times7$ input convolution, followed by max-pooling with the stride of $2$, and pool size of $3\times3$, is used before identity and convolutional blocks. 
By stacking these blocks on top of each other (see Fig.~\ref{fig:network}), an encoder-$1$ has formed to get the feature map, where the notation ($n\times $) under the identity block denotes the number of repetitions ($n$ times).
The different blocks, of encoder-$1$ ($E_{1n}$ and $n=1, 2, ..., 5$), downsample the input image resolutions in half of the input resolutions, while the resolution inside the blocks is kept constant. The outputs of those blocks generate the feature maps with different scales.  
Within the encoder-$2$, three components, of information flow blocks, are used, which were initially proposed by \citet{chollet2017xception}, such as entry flow, middle flow, and exit flow, as depicted in Fig.~\ref{fig:network}.
The batch, of input images, firstly passes through the input flow, then the central flow, eight times ($8\times $) repeated, and finally through the exit flow.
All flows, as in the proposed network (see in Fig.~\ref{fig:network}), have Depth-wise Separable Convolution (DwSC) \citep{chollet2017xception} and residual connections. 
The former one has used to create a lightweight network, while the latter has the advantages discussed earlier.
Thus, by utilizing two different types of encoders, we can learn two types of feature maps from the same input images. 
However, the different blocks of encoder-$2$ ($E_{2n}$ and $n=1, 2, ..., 5$), downsample the input image resolutions in half of the input resolutions, while the resolution inside the blocks is kept constant. The outputs of those blocks also generate feature maps with different scales.  
However, two different $2$D feature maps of different encoders are concatenated, in channel-wise, to enhance the depth information of the feature map. We use differently scaled feature maps to build the proposed CVR-Net, where each feature map is passed through the Fully Connected Layer (FCL) block. A Global Average Pooling (GAP) \citep{lin2013network} layer and four fully connected layers are used in our FCL block, where the GAP layer performs an extreme dimensionality reduction to avoid overfitting. An $height \times weight \times depth$ dimensional tensor, in GAP, is reduced to a $1 \times 1 \times depth$ vector by transferring $height \times width$ feature map to a single number, which contributes to the lightweight design of the proposed CVR-Net. Table~\ref{tab:models} presents the implementational details of the proposed CVR-Net.  
\begin{table*}[!ht]
\caption{Details of the proposed CVR-Net have used feature maps, shapes, and the number of parameters, where the input resolution is $M\times N$ pixels.}
\footnotesize
\begin{tabular}{cccc}
\hline
\rowcolor[HTML]{C0C0C0} 
Feature block & Shape of features & Prediction & Parameters \\ \hline
$E_{13}$  &   $\frac{M}{8} \times \frac{N}{8} \times 512$ &  $P_1=FCL(E_{13})$   &   \begin{tabular}[c]{@{}l@{}}  $1,796,867$  \end{tabular}          \\ \hline
$E_{14}$       &   $\frac{M}{16} \times \frac{N}{16} \times 1024$  & $P_2=FCL(E_{14})$  &    \begin{tabular}[c]{@{}l@{}}  $9,181,827$  \end{tabular}         \\ \hline
$[E_{15} \concat E_{25} ]$        &  $\frac{M}{32} \times \frac{N}{32} \times 4096$     &  $P_3=FCL([E_{15} \concat E_{25} ])$  &   \begin{tabular}[c]{@{}l@{}}  $46, 620, 971$  \end{tabular}        \\ \hline
$E_{23}$       &  $\frac{M}{8} \times \frac{N}{8} \times 512$     & $P_4=FCL(E_{23})$  &     \begin{tabular}[c]{@{}l@{}}  $1,371,131$  \end{tabular}       \\ \hline
$E_{24}$       &   $\frac{M}{16} \times \frac{N}{16} \times 1024$    &  $P_5=FCL(E_{24})$   & \begin{tabular}[c]{@{}l@{}} $15,954,283$  \end{tabular}        \\ \hline
\multicolumn{2}{c}{\textbf{Proposed CVR-Net}} & $P=Avg(P_1\sim P_5)$ & $48,596,087$  \\ \hline
\end{tabular}
\label{tab:models}
\end{table*}
We utilize the feature maps $E_{13}\sim E_{15}$ from encoder-$1$ and $E_{23}\sim E_{25}$ from encoder-$2$, where we concatenate $E_{15}$ and $E_{25}$ to increase the depth of the feature information. The final prediction, in CVR-Net, is the average of different probabilities, such as $P_1$, $P_2$, $P_3$, $P_4$, and $P_5$ respectively for $E_{13}$, $E_{14}$, $[E_{15} \concat E_{25} ]$, $E_{23}$, and $E_{24}$, which was trained end-to-end fashion.
However, designing of such a multi-encoder and multi-scale network, as CVR-Net, has several benefits, especially for the small datasets, such as: if one encoder fails to generate responsible features, another encoder can compensate it and vice-versa; if the feature quality is reduced in the deeper blocks (lower resolution), the prior blocks (higher resolution) can also compensate it and vice-versa; if one or more $P$ predicts wrong class, other $P$ can overcome it, as the final result is average of all $P$'s. Another positive prospect of the CVR-Net is that during the training, it can be anticipated that if the gradient of one or more branches vanishes, then other branches can recover it as the final gradient is the average of all the individual gradient.


\subsection{Training Protocol and Evaluation}
\label{Training}
\paragraph{\textbf{Training Protocol}}
The preprocessing of the input images is the crucial requirement of the deep CNN models, as images serve as fuel in the learning of those models. However, as a preprocessing, we apply image augmentations, class rebalancing, and resizing. The supervised learning systems, in medical imaging domains, suffer from the limited size of the datasets, which is one reason for the overfitted model. 
The data augmentations can partly overcome such overfitting, where augmentations utilize either data warping or oversampling for augmenting the training dataset synthetically \citep{shorten2019survey}.  
In this research, we apply different geometric transformations, as an augmentation, such as rotation, height \& width shift, and horizontal \& vertical flipping. 
Imbalanced class distribution is another common phenomenon in the medical imaging domain, where the positive class is underrepresented compared to other classes, as the manual annotation is expensive. However, to alleviate this problem, we penalize the majority class by weighting the loss function, where such a weighting pays more attention to samples from the underrepresented class. We estimate the weights of each class by $W_j = N_j/N$, where $W_j$, $N$, and $N_j$ are the weight for class $j$, the total number of samples, and the number of samples in class $j$, respectively. 
As we noticed that most of the images of all the datasets have a $1:1$ aspect ratio, all images were resized to $224\times 224$ pixels using nearest-neighbor interpolation.
Additionally, the scarcity of relatively small medical image datasets has been partially overcome by employing a transfer learning \citep{shin2016deep,tajbakhsh2016convolutional} where the previously trained model, so-called pre-trained model, is used to initialize the kernels rather than random initialization. We utilize the ImageNet \citep{krizhevsky2012imagenet} pre-trained weights to initialize the kernels of the proposed CVR-Net.  
We employ categorical cross-entropy as a loss function and accuracy as a metric for training our CVR-Net for all the datasets. The  loss  function has been optimized using the Adam \citep{kingma2014adam} optimizer with initial learning rate ($LR$), exponential decay rates ($\beta{_1},\,\beta{_2}$) as $LR=0.0001$, $\beta{_1}=0.9$, and $\beta{_2}=0.999$ respectively without AMSGrad variant. 
The initial learning rate is reduced after $12$ epochs by $10.0\,\%$ if validation loss stops improving.


\paragraph{\textbf{Evaluation}}
We use different metrics, such as recall, precision, F1-score, and accuracy, to evaluate our multi-tasking CVR-Net for coronavirus recognition, which are mathematically defined as follows: 
\[Recall = \frac{TP}{TP+FN}\]

\[Precision = \frac{TP}{TP+FP}\]

\[F1-score = \frac{2 \times TP}{2\times TP+FN+FP}\]

\[Accuracy = \frac{TP+TN}{TP+FN+FP+TN}\]

\noindent where the TP, FN, FP, and TN respectively denote true positive (patient with coronavirus symptoms recognized as the positive patient), false negative (patient with coronavirus symptoms recognized as the negative patient), false positive (patient without coronavirus symptoms recognized as the positive patient), and true negative (patient without coronavirus symptoms recognized as the negative patient). 
The recall quantifies the type-II error (the patient, with the positive syndromes, inappropriately fails to be nullified), and precision quantifies the positive predictive values (percentage of truly positive recognition among all the positive recognition). 
The F$1$-score indicates the harmonic mean of recall and precision, which shows the trade-off between them. Accuracy quantifies the fraction of correct predictions (both positive and negative).


\section{Experiments and Results}
\label{ResultsandDiscussion}
At the beginning of this section, we present the quantitative results for coronavirus recognition, applying the proposed CVR-Net with different datasets having a different number of classes. Finally, in the end, we compare our multi-tasking results with several recent state-of-the-art results, on the same datasets, to validate our proposal.

Table~\ref{tab:allresults} presents all the quantitative results, for the coronavirus recognition, utilizing different datasets and our proposed CVR-Net. The evaluation metrics, for each fold of each task, with average values are reported, in Table~\ref{tab:allresults}, for evaluating the inter-fold variations.    
\begin{table*}[!ht]
\caption{Coronavirus recognition results, applying the proposed CVR-Net, from different comprehensive experiments on the different datasets with different tasks. }
\footnotesize
\begin{tabular}{ccccccc}
\hline
\rowcolor[HTML]{C0C0C0} 
\cellcolor[HTML]{C0C0C0}                           & \cellcolor[HTML]{C0C0C0}                             & \cellcolor[HTML]{C0C0C0}                        & \multicolumn{4}{c}{\cellcolor[HTML]{C0C0C0}Metrics} \\ \cline{4-7} 
\rowcolor[HTML]{C0C0C0} 
\multirow{-2}{*}{\cellcolor[HTML]{C0C0C0}Datasets} & \multirow{-2}{*}{\cellcolor[HTML]{C0C0C0}Task types} & \multirow{-2}{*}{\cellcolor[HTML]{C0C0C0}Folds} & Recall    & Precision    & F1-score    & Accuracy   \\ \hline
                                                   &                                                      & Fold-1                                          &  $0.995$         &     $0.995$         &  $0.995$           &   $0.998$         \\
                                                   &                                                      & Fold-2                                          &  $0.998$          &       $0.998$        &        $0.998$      &     $0.998$        \\
                                                   &                                                      & Fold-3                                          &  $0.996$          &  $0.996$             &  $0.996$            &  $0.996$           \\
                                                   &                                                      & Fold-4                                          &   $0.998$        &    $0.998$          &  $0.998$           &   $0.998$         \\
                                                   &                                                      & Fold-5                                          &   $0.998$         &  $0.998$             &   $0.998$           &    $0.998$         \\ \cline{3-7} 
                                                   & \multirow{-6}{*}{Task-1: 2-class}                    & \textbf{Average}                                         & $0.997\pm 0.001$           & $0.997\pm 0.001$              & $0.997\pm 0.001$            &  $0.998\pm 0.001$          \\ \cline{2-7} 
                                                   &                                                      & Fold-1                                          &  $0.964$         &    $0.964$          &      $0.963$       &  $0.964$          \\
                                                   &                                                      & Fold-2                                          &       $0.966$    &    $0.966$          &       $0.966$      &    $0.966$        \\
                                                   &                                                      & Fold-3                                          &    $0.955$       &        $0.955$      &     $0.955$        & $0.955$           \\
                                                   &                                                      & Fold-4                                         &  $0.969$        &    $0.968$            &    $0.968$          &  $0.969$          \\
                                                   &                                                      & Fold-5                                          &     $0.965$      &    $0.964$          &  $0.964$           & $0.965$           \\ \cline{3-7} 
                                                   & \multirow{-6}{*}{Task-2: 3-class}                    & \textbf{Average}                                         & $	0.964\pm 0.005$           & $0.963\pm 0.004 $          & $0.963\pm 0.004$            &  $0.964\pm 0.005$          \\ \cline{2-7} 
                                                   &                                                      & Fold-1                                          &   0.811        &    $0.808$          &   $0.808$          &    $0.811$        \\
                                                   &                                                      & Fold-2                                          &    $0.818$       &     $0.813$         &  $0.813$           &  $0.818$         \\
                                                   &                                                      & Fold-3                                          &   $0.823$        &   $0.819$           & $0.819$            &    $0.823$        \\
                                                   &                                                      & Fold-4                                          &   $0.817$        &  $0.815$            &      $0.815$       &  $0.817$          \\
                                                   &                                                      & Fold-5                                          &  $0.831$         &    $0.827$          &   $0.826$          &      $0.831$     \\ \cline{3-7} 
\multirow{-18}{*}{Dataset-1}                       & \multirow{-6}{*}{Task-3: 4-class}                    & \textbf{Average}                                         &  $0.820\pm 0.007$          &   $0.816\pm 0.006$           &  $0.816\pm 0.006$           &   $0.820\pm 0.007$         \\ \hline
                                                   &                                                      & Fold-1                                         &    $0.972$                & $0.972$          &           $0.972$  &  $0.972$          \\
                                                   &                                                      & Fold-2                                          &            $0.960$  &     $0.961$      &     $0.960$        &     $0.960$       \\
                                                   &                                                      & Fold-3                                          &  $0.965$            & $0.965$          & $0.965$             &       $0.965$     \\
                                                   &                                                      & Fold-4                                          &            $0.952$  & $0.952$          &   $0.951$          &     $0.952$       \\
                                                   &                                                      & Fold-5                                          &  $0.955$             &     $0.955$      &     $0.955$        &    $0.955$        \\ \cline{3-7} 
\multirow{-6}{*}{Dataset-3}                        & \multirow{-6}{*}{Task-4: 3-class}                    & \textbf{Average}                                         &  $0.961\pm 0.007$            &   $	0.961\pm 0.007$        & $0.961\pm 0.007$             &  $0.961\pm 0.007$        \\ \hline
Dataset-5                        & {Task-5: 2-class}                    & -                                         &     $0.780$         &     $0.780$      &  $0.780$           &     $0.780$       \\ \hline
\end{tabular}
\label{tab:allresults}
\end{table*}


\paragraph{\textbf{Experiment-$1$}}
We have trained and evaluated the proposed CVR-Net on binary classification problem (Task-$1$), with $5856$-negative (NOR) and $500$-positive (NCP) images, applying $5$-fold cross-validations. For Task-$1$, the proposed CVR-Net achieved an average accuracy of $99.8\,\%$ on dataset-$1$, while the average recall, precision, and F1-score are $99.7\,\%$, $99.7\,\%$, and $99.7\,\%$, respectively. The binary results, as presented in Table~\ref{tab:allresults}, demonstrate that $99.7\,\%$ NCP images are correctly recognized as NCP, while the positive predictive value is also $99.7\,\%$. The inter-fold variations, for all the metrics, are also as small as $0.1\,\%$, which discloses commendable robustness of the proposed CVR-Net. 
The details class-wise results, from the proposed CVR-Net, are presented in a confusion matrix in Table~\ref{tab:cm1}. 
\begin{table}[!ht]
\footnotesize
\centering
\caption{Confusion matrix of Task-$1$ ($2$-class) of Dataset-$1$ employing the proposed CVR-Net, where the samples are the summation from each fold.}
\begin{tabular}{ccclcl}
\hline
\multicolumn{2}{c}{}                   & \multicolumn{4}{c}{Actual} 
                                                \\ \cline{3-6} 
\multicolumn{2}{c}{\multirow{-2}{*}} & \multicolumn{2}{c}{NOR}                                              & \multicolumn{2}{c}{NCP}                                            \\ \hline
                                           & NOR               & \multicolumn{2}{c}{\cellcolor[HTML]{C0C0C0}\begin{tabular}[c]{@{}c@{}}$5848$ \\ $99.86\,\%$\end{tabular}} & \multicolumn{2}{c}{\begin{tabular}[c]{@{}c@{}}$7$ \\ $1.40\,\%$\end{tabular}}                         \\ \cline{2-6} 
\multirow{-2}{*}{Predicted}                & NCP               & 

\multicolumn{2}{c}{\begin{tabular}[c]{@{}c@{}}$8$ \\ $0.14\,\%$\end{tabular}}                           & \multicolumn{2}{c}{\cellcolor[HTML]{C0C0C0}\begin{tabular}[c]{@{}c@{}}$493$ \\ $98.60\,\%$\end{tabular}} \\ \hline
\end{tabular}
\label{tab:cm1}
\end{table}
It exposes that out of $500$-NCP samples, the proposed model successfully can recognize $493$ samples as NCP, while only $7$ samples are predicted as NOR (false negative). 
Table~\ref{tab:cm1} also reveals that the power of a binary test (probability of rejecting the null hypothesis) is $98.6\,\%$, which is an excellent outcome, on the utilized dataset, by the proposed CVR-Net.


\paragraph{\textbf{Experiment-$2$}}
We split the NOR images into NOR and pneumonia (CPN) classes, and then, we train and evaluate the proposed model on $3$-class problem (Task-2), where we have $1583$-NOR, $4273$-CPN, and $500$-NCP images, applying $5$-fold cross-validations. For this task, the obtained accuracy, recall, precision, and F1-score are $96.4\,\%$, $96.3\,\%$, $96.3\,\%$, and $96.4\,\%$, respectively (see in Table~\ref{tab:allresults}). Those results confess that on an average $96.4\,\%$-samples are correctly recognized by the proposed CVR-Net with type-II error and positive predictive value of $3.6\,\%$, and $96.3\,\%$, respectively. It is also noteworthy, from Table~\ref{tab:allresults}, that the inter-fold variation is increasing with the decreased performances for all the metrics.
The details class-wise results, from the proposed CVR-Net for this task, are presented in a confusion matrix in Table~\ref{tab:cm2}.  
\begin{table}[!ht]
\footnotesize
\centering
\caption{Confusion matrix of Task-$2$ ($3$-class) of Dataset-$1$ employing the proposed CVR-Net, where the samples are the summation from each fold.}
\begin{tabular}{ccccc}
\hline
                            &     & \multicolumn{3}{c}{Actual}                                                     \\ \cline{3-5} 
                            &     & NOR                      & CPN                       & NCP                      \\ \hline
                            & NOR & \cellcolor[HTML]{C0C0C0} \begin{tabular}[c]{@{}c@{}}1461 \\ $92.29\,\%$ \end{tabular} &    \begin{tabular}[c]{@{}c@{}}80 \\ $1.87\,\%$ \end{tabular}                       &   \begin{tabular}[c]{@{}c@{}}6 \\ $1.20\,\%$ \end{tabular}                        \\ \cline{2-5} 
                            & CPN  &      \begin{tabular}[c]{@{}c@{}}119 \\ $7.52\,\%$ \end{tabular}                     & \cellcolor[HTML]{C0C0C0} \begin{tabular}[c]{@{}c@{}}4185 \\ $97.94\,\%$ \end{tabular} &   \begin{tabular}[c]{@{}c@{}}15 \\ $3.00\,\%$ \end{tabular}                        \\ \cline{2-5} 
\multirow{-3}{*}{Predicted} & NCP &     \begin{tabular}[c]{@{}c@{}}3 \\ $0.19\,\%$ \end{tabular}                      &        \begin{tabular}[c]{@{}c@{}}8 \\ $0.19\,\%$ \end{tabular}                   & \cellcolor[HTML]{C0C0C0} \begin{tabular}[c]{@{}c@{}}479 \\ $95.8\,\%$ \end{tabular} \\ \hline
\end{tabular}
\label{tab:cm2}
\end{table}
It shows that the addition of new class (CPN) with earlier two classes (NOR and NCP), as in Task-$1$, reduces the recognition rate of coronavirus from $98.6\,\%$ to $95.8\,\%$, where Task-$2$ has $21$-false negatives out of $500$ samples. It is also observable that out of $21$-false negatives, $15$-NCP samples are recognized as CPN, which affirms that there is a high degree of similarity between CPN and NCP classes. Moreover, $80$-CPN and $119$-NOR are recognized as the NOR and CPN, respectively. Which again reveals the inter-class similarity between CPN and NOR classes.


\paragraph{\textbf{Experiment-$3$}}
We further break the CPN class into pneumonia bacterial (CPB) and pneumonia viral (CPV) classes. Thus, we have a total of $4$ classes in Task-$3$, where it has $1583$-NOR, $2780$-CPB, $1493$-CPV, and $500$-NCP images. We also apply $5$-fold cross-validation in this experiment. The proposed CVR-Net, for this experiment (Task-3), produces the recognition results with accuracy, recall, precision, and F1-score of $82.0\,\%$, $82.0\,\%$, $81.6\,\%$, and $81.6\,\%$, respectively, with increased standard deviation comparing two previous experiments (see Table~\ref{tab:allresults}). Those results, in this Task-$3$, expose that the CVR-Net recognizes the classes with higher error rates than two previous experiments, where it has a false-negative rate and positive predictive value of $18.0\,\%$ and $81.6\,\%$, respectively. 
It is also noteworthy that the positive predictive value and type-II error have been reduced by the margins of $18.2\,\%$ and $17.8\,\%$, respectively, than binary Task-$1$ (see in Table~\ref{tab:allresults}), which indicates that additional $17.8\,\%$-NCP samples are recognized as other classes (false negative) in Task-$3$. However, a further class-wise investigation is given in a confusion matrix, as shown in Table~\ref{tab:cm3}, which exhibits that the CVR-Net fails to recognize the coronavirus in $40$ cases. 
\begin{table}[!ht]
\caption{Confusion matrix of Task-$3$ ($4$-class) of Dataset-$1$ employing the proposed CVR-Net, where the samples are the summation from each fold.}
\footnotesize
\begin{tabular}{clcccc}
\hline
\multicolumn{2}{l}{}                   & \multicolumn{4}{c}{Actual}                                                                                                                                                                                                                                                                    \\ \cline{3-6} 
\multicolumn{2}{l}{\multirow{-2}{*}{}} & \multicolumn{1}{l}{NOR}                                               & \multicolumn{1}{l}{CPB}                                               & \multicolumn{1}{l}{CPV}                                               & \multicolumn{1}{l}{NCP}                                               \\ \hline
                              & NOR    & \cellcolor[HTML]{C0C0C0}\begin{tabular}[c]{@{}c@{}}1498\\ $94.63\,\%$\end{tabular} & \begin{tabular}[c]{@{}c@{}}76\\ $2.73\,\%$\end{tabular}                         & \begin{tabular}[c]{@{}c@{}}74\\ $4.96\,\%$\end{tabular}                         & \begin{tabular}[c]{@{}c@{}}13\\ $2.60\,\%$\end{tabular}                         \\ \cline{2-6} 
                              & CPB    & \begin{tabular}[c]{@{}c@{}}31\\ $1.96\,\%$\end{tabular}                         & \cellcolor[HTML]{C0C0C0}\begin{tabular}[c]{@{}c@{}}2369\\ $85.22\,\%$\end{tabular} & \begin{tabular}[c]{@{}c@{}}529\\ $35.43\,\%$\end{tabular}                         & \begin{tabular}[c]{@{}c@{}}14\\ $2.80\,\%$\end{tabular}                         \\ \cline{2-6} 
                              & CPV    & \begin{tabular}[c]{@{}c@{}}52\\ $3.29\,\%$\end{tabular}                         & \begin{tabular}[c]{@{}c@{}}330\\ $11.87\,\%$\end{tabular}                         & \cellcolor[HTML]{C0C0C0}\begin{tabular}[c]{@{}c@{}}885\\ $59.28\,\%$\end{tabular} & \begin{tabular}[c]{@{}c@{}}13\\ $2.60\,\%$\end{tabular}                         \\ \cline{2-6} 
\multirow{-4}{*}{Predict}     & NCP    & \begin{tabular}[c]{@{}c@{}}2\\ $0.12\,\%$\end{tabular}                         & \begin{tabular}[c]{@{}c@{}}5\\ $0.18\,\%$\end{tabular}                         & \begin{tabular}[c]{@{}c@{}}5\\ $0.33\,\%$\end{tabular}                         & \cellcolor[HTML]{C0C0C0}\begin{tabular}[c]{@{}c@{}}460\\ $92.00\,\%$\end{tabular} \\ \hline
\end{tabular}
\label{tab:cm3}
\end{table}
Table~\ref{tab:cm3} shows that the proposed CVR-Net can successfully recognize $460$-NCP samples as NCP, where it erroneously predict $27$ samples as CPB \& CPV and $13$ samples as NOR. It is also noteworthy that $35.43\,\%$-CPV, and $11.87\,\%$-CPB are respectively recognized as CPB and CPV.


\paragraph{\textbf{Experiment-$4$ \& Experiment-$5$}}
In these two experiments, to train and evaluate the proposed CVR-Net, we use dataset-$2$, with $3$-classes (Task-$4$), and dataset-$3$, with $2$-classes (Task-$5$), as presented in Table~\ref{tab:dataset}.
Table~\ref{tab:allresults} shows that our model has accuracy, recall, precision, and F$1$-score of $96.1\,\%$, $96.1\,\%$, $96.1\,\%$, and $96.1\,\%$, respectively, for dataset-$2$ (Task-$4$), which are $78.0\,\%$, $78.0\,\%$, $78.0\,\%$, and $78.0\,\%$, respectively, for dataset-$3$ (Task-$5$). 
Those results on dataset-$2$ and dataset-$3$ demonstrate that our CVR-Net model recognizes the coronavirus with type-II errors as $3.9\,\%$, and $22.0\,\%$, respectively.  
Table~\ref{tab:cm4} and Table~\ref{tab:cm5} show the confusion matrix from our proposed CVR-Net utilizing the dataset-$2$ and dataset-$3$, respectively.       
\begin{table}[!ht]
\footnotesize
\centering
\caption{Confusion matrix of Task-$4$ ($3$-class) of Dataset-$2$ employing the proposed CVR-Net, where the samples are the summation from each fold.}
\begin{tabular}{ccccc}
\hline
                            &     & \multicolumn{3}{c}{Actual}                                                     \\ \cline{3-5} 
                            &     & NOR                      & CPN                       & NCP                      \\ \hline
                            & NOR & \cellcolor[HTML]{C0C0C0} \begin{tabular}[c]{@{}c@{}}1522 \\ $92.35\,\%$ \end{tabular} &    \begin{tabular}[c]{@{}c@{}}88 \\ $2.01\,\%$ \end{tabular}                       &   \begin{tabular}[c]{@{}c@{}}7 \\ $1.40\,\%$ \end{tabular}                        \\ \cline{2-5} 
                            & CPN  &      \begin{tabular}[c]{@{}c@{}}126 \\ $7.65\,\%$ \end{tabular}                     & \cellcolor[HTML]{C0C0C0} \begin{tabular}[c]{@{}c@{}}4276 \\ $97.83\,\%$ \end{tabular} &   \begin{tabular}[c]{@{}c@{}}26 \\ $5.20\,\%$ \end{tabular}                        \\ \cline{2-5} 
\multirow{-3}{*}{Predicted} & NCP &     \begin{tabular}[c]{@{}c@{}}0 \\ $0.00\,\%$ \end{tabular}                      &        \begin{tabular}[c]{@{}c@{}}7 \\ $0.16\,\%$ \end{tabular}                   & \cellcolor[HTML]{C0C0C0} \begin{tabular}[c]{@{}c@{}}467 \\ $93.4\,\%$ \end{tabular} \\ \hline
\end{tabular}
\label{tab:cm4}
\end{table}
\begin{table}[!ht]
\footnotesize
\centering
\caption{Confusion matrix of Task-$5$ ($2$-class) of Dataset-$3$ employing the proposed CVR-Net, where the samples are the summation from each fold.}
\begin{tabular}{ccclcl}
\hline

\multicolumn{2}{c}{}                   & \multicolumn{4}{c}{Actual} 
                                                \\ \cline{3-6} 
\multicolumn{2}{c}{\multirow{-2}{*}} & \multicolumn{2}{c}{NOR}                                              & \multicolumn{2}{c}{NCP}                                            \\ \hline
                                           & NOR               & \multicolumn{2}{c}{\cellcolor[HTML]{C0C0C0}\begin{tabular}[c]{@{}c@{}}$87$ \\ $82.86\,\%$\end{tabular}} & \multicolumn{2}{c}{\begin{tabular}[c]{@{}c@{}}$26$ \\ $26.53\,\%$\end{tabular}}                         \\ \cline{2-6}

\multirow{-2}{*}{Predicted}                & NCP               & 

\multicolumn{2}{c}{\begin{tabular}[c]{@{}c@{}}$18$ \\ $17.14\,\%$\end{tabular}}                           & \multicolumn{2}{c}{\cellcolor[HTML]{C0C0C0}\begin{tabular}[c]{@{}c@{}}$72$ \\ $73.47\,\%$\end{tabular}} \\ \hline
\end{tabular}
\label{tab:cm5}
\end{table}
The matrix, as shown in Table~\ref{tab:cm4}, reveals the FN and FP for coronavirus recognition, where the number of wrongly classified images (type-I or type-II errors) is $126/1648\,(7.65\,\%)$, $95/4371\,(2.17\,\%)$, and $33/500\,(6.60\,\%)$ respectively for the NOR, CPN, and NCP. In total, $6263$ images (out of $6519$) are successfully recognized as their respective classes, especially $467$ images (out of $500$) for coronavirus, which exhibits the praiseworthy success of the proposed CVR-Net for the correct recognition of coronavirus.     
Again, the confusion matrix, as presented in Table~\ref{tab:cm5}, shows that $82.86\,\%$-NOR samples are correctly classified as NOR, whereas $17.14\,\%$-NOR samples are wrongly classified as NCP. 
On the other hand,  $73.47\,\%$-NCP samples are correctly classified as NCP, whereas $26.53\,\%$-NCP samples are wrongly classified as NOR. 
Although the performance of the CVR-Net on dataset-$3$ is not as high as in the dataset-$1$ and dataset-$2$ (see in Table~\ref{tab:allresults}), it is still better as the utilized dataset is very small in size comparing other two datasets (see in Table~\ref{tab:dataset}).


\paragraph{\textbf{Results Comparison}}
Table~\ref{tab:previousWork2016} represents the performance comparison of the proposed CVR-Net with other recent state-of-the-art methods with dataset-$1$ (Task-$1$ and Task-$2$) and dataset-$3$ (Task-$5$). The remaining two other tasks, (Task-$3$ and Task-$4$) are not reported in Table~\ref{tab:previousWork2016}, as state-of-the-art methods were not trained and tested on these datasets.   
To enhance the recognition performance, authors, in several new methods, utilized more the external data to train their networks, which are not publicly available yet.
The improvement of the recognition network may not be due to the superiority of the network itself, but the characteristics of the external data, similar to the test datasets.
However, we have reported the results of the methods, which were trained and tested on the same datasets, for fairness in comparison.  
The proposed CVR-Net produces the best recognition results, as presented in Table~\ref{tab:previousWork2016}, for seven out of the nine cases while performing second-best with the winning methods on the other two cases. 
\begin{table*}[!ht]
\caption{The state-of-the-art comparison with proposed CVR-Net, for three different tasks, was trained, validated, and tested on the same dataset. The best, second-best and third-best metrics are denoted by bold font, underline, and double underline.}
\tiny
\begin{tabular}{l|ccc|ccc|ccc}
\hline
\rowcolor[HTML]{C0C0C0} 
\cellcolor[HTML]{C0C0C0}                          & \multicolumn{3}{c|}{\cellcolor[HTML]{C0C0C0}Task-1} & \multicolumn{3}{c|}{\cellcolor[HTML]{C0C0C0}Task-2} & 
\multicolumn{3}{c}{\cellcolor[HTML]{C0C0C0}Task-5}\\ \cline{2-10} 
\rowcolor[HTML]{C0C0C0} 
\multirow{-2}{*}{\cellcolor[HTML]{C0C0C0}Methods} & Re           & Pr           & Ac          & Re           & Pr           & Ac         & Re           & Pr           & Ac       \\ \hline

VGG-$19$ \citep{apostolopoulos2020covid} &  - &    - &     $0.987$     &  -            &    -    &     $0.935$         & -        &            -                                    &     -   \\ \hline 

Xception \citep{apostolopoulos2020covid} &  - &    - &     $0.856$     &  -            &    -    &     $0.928$         & -        &            -     &     -      \\ \hline 

Covid-Net \citep{wang2020covid} &  - &    - &    -     &  $\doubleunderline{0.933}$            &    $\doubleunderline{0.937}$   &     $\doubleunderline{0.933}$         & -        &            -     &     -     \\ \hline 

ResNet-$50$ \citep{sethy2020detection} &  $\doubleunderline{0.973}$ & - &    $0.954$    &  -   &    -   &     -         & -        &            -     &     -             \\ \hline

VGG-$19$ \citep{hemdan2020covidx} &  $0.900$ &   $0.915$ &    $0.900$    &  -   &    -   &     -         & -        &            -     &       \\ \hline

ResNet-$50$ \citep{narin2020automatic} &  $0.960$ &   $\textbf{1.0}$ &    $0.980$    &  -   &    -   &     -         & -        &            -     &      \\ \hline

InceptionV$3$ \citep{narin2020automatic} &  $0.940$ &   $\textbf{1.0}$ &    $0.970$    &  -   &    -   &     -         & -        &            -     &     -        \\ \hline

DarkNet \citep{ozturk2020automated} &  $0.951$ &   $0.980$ &    $\doubleunderline{0.981}$    &  $0.854$   &    $0.900$   &     $0.870$        & -        &            -     &       \\ \hline

CoroNet(Xception) \citep{khan2020coronet} &  $\underline{0.993}$ &   $\doubleunderline{0.983}$ &    $\underline{0.990}$    &  $\textbf{0.969}$   & $\underline{0.950}$      &     $\underline{0.950}$        &  -      &   -     &  -    \\ \hline

VGG-$16$ \citep{he2020sample}   & -      &   -    &    -  &  -     &  - & -  &  - &   - &    $\doubleunderline{0.760}$ \\ \hline 
ResNet-$18$ \citep{he2020sample} & -      &   -    &    -  &  -     &  - & -  &  - &   - &    $0.740$ \\ \hline 
EfficientNet-b0 \citep{he2020sample}    & -      &   -    &    -  &  -     &  - & -  &  - &   - &    $\underline{0.770}$ \\ \hline 
CRNet \citep{he2020sample}   & -      &   -    &    -  &  -     &  - & -  &  - &   - &    $0.730$ \\ \hline 

\textbf{CVR-Net (Proposed, 2020)} &  $\textbf{0.997}$ &   $\underline{0.997}$ &    $\textbf{0.998}$    &  $\underline{0.964}$   &    $\textbf{0.963}$   &    $\textbf{0.964}$   &  $\textbf{0.780}$ &   $\textbf{0.780}$ &    $\textbf{0.780}$ \\ \hline 
\multicolumn{10}{l}{\scriptsize{Re: Recall, Pr: Precision, and Ac: Accuracy.}}
\end{tabular}
\label{tab:previousWork2016}
\end{table*}
Firstly, the proposed CVR-Net yields the best results, for Task-$1$, concerning the accuracy and type-II errors (recall) by beating the second-best state-of-the-art CoroNet \citep{khan2020coronet} with the margins of $0.8\,\%$ and $0.4\,\%$, respectively.
Concerning the positive predictive value, CVR-Net, for Task-$1$, is behind the state-of-the-art ResNet-$50$ \& InceptionV$3$ \citep{narin2020automatic} by $0.3\,\%$, but it outperforms the third-best CoroNet \citep{khan2020coronet} by a $1.4\,\%$ margin with respect to the same metric. 
Secondly, for Task-$2$, CVR-Net beats the second-best CoroNet \citep{khan2020coronet} by the margins of $1.3\,\%$, and $1.4\,\%$ respectively for precision and accuracy. Although, for the same task and type-II errors (recall), our CVR-Net is behind the winner CoroNet \citep{khan2020coronet} by $0.5\,\%$, it outperforms the third-best recall of COVID-Net \citep{wang2020covid} by a margin of $3.1\,\%$.
Thirdly, for Task-$5$ and grand challenge dataset, our proposed CVR-Net outperforms all the methods, such as VGG-$16$ \citep{he2020sample}, ResNet-$18$ \citep{he2020sample}, EfficientNet-b0 \citep{he2020sample}, and CRNet \citep{he2020sample}, for all the metrics (see in Table~\ref{tab:previousWork2016}). The above discussions show that the performance of the proposed CVR-Net for coronavirus recognition is praiseworthy for all the utilized datasets.


\section{Discussion}
\label{Discussion}
The COVID-$19$ pandemic has a disastrous effect on the health and well-being of the global population. Effective and early screening of infected patients is a critical and crucial step to fight against the COVID-$19$ epidemic, where the examination and investigation of chest radiography images, via any CAD tool, is one of the vital screening approaches. Recent studies on COVID-$19$ patients show that there are several coronavirus characteristics in chest radiography images. Motivated by this and inspired by the research community's open-source endeavors, we aimed to design an automated image classifier for the recognition of the coronavirus utilizing the chest radiography images.

However, to design such a classifier, CNNs are better-choice as they automatically learn low-, middle-, and high-level features directly from the input images. Finally, fully-connected neural networks, also known as multilayer perceptron, classify those features. However, such CNN-based classifiers' training is an arduously challenging process, especially when the training is with a smaller dataset as in the COVID-$19$ datasets. 
There are several commonly occurring limitations in current CNN-based classifiers; it is prone to overfitting, vanishing gradient problem, and amount of the network's depth with the following number of times of subsampling.
However, in this article, we proposed an end-to-end network called CVR-Net for automated coronavirus prediction by considering the limitations mentioned earlier in network design. In the proposed CVR-Net the aggregation, of the different encoders and their different scales partially alleviates those limitations, as if one or more members of the ensembled CVR-Net, fail to predict other can compensate it. The final cost function can not be zero, as it is the summation of each cost. Thus, the gradient is always non-zero in our proposed CVR-Net.

The class-wise results, for all the tasks (Task-$1$ to Task-$5$), in Table~\ref{tab:cm1} to Table~\ref{tab:cm5}, experimentally show that the metrics for all classes are similar for each task, although imbalanced data distributions are utilized in our model.
The positive NCP class is highly underrepresented, for all the tasks (see in Table~\ref{tab:dataset}), still it results compatible with other classes. However, the employment of class rebalancing by penalizing the cost of the overrepresented class, in this article (see in subsection \ref{Training}), is the crucial reason behind these balanced performances by the proposed CVR-Net. The results for all the tasks, as in Table~\ref{tab:allresults}, exhibit that the inter-fold variation is very less, which ensures the better-robustness of the CVR-Net for coronavirus and pneumonia recognition. The multi-scale-multi-encoder ensemble in CVR-Net and appliance of the reasonable image augmentations and transferring the weights from the ImageNet as a preprocessing, during the training of CVR-Net, are the noteworthy catalyst of obtaining the robust recognition results.

\begin{figure*}[!ht]
  \centering
  \subfloat{\includegraphics[width=13cm, height= 6.5cm]{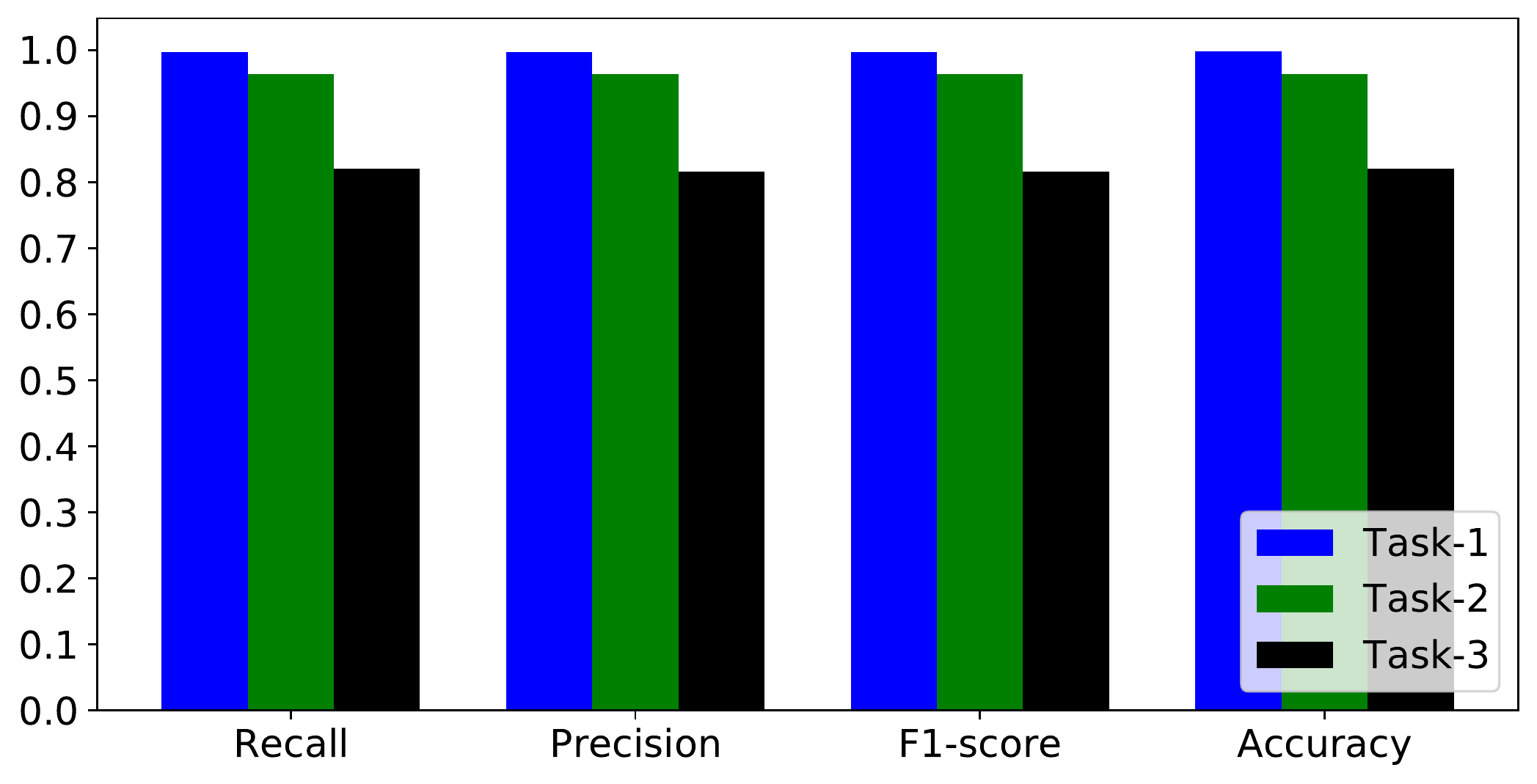}} 
  \caption{Impact of increasing the number of classes with the same number of total training samples, where bars with blue, green, and black colors denote the CVR-Net results for $2$-class, $3$-class, and 4-class, respectively.}
  \label{fig:classwiseperfm}
\end{figure*}

The experimental results on dataset-$1$ from the CVR-Net, in Table~\ref{tab:allresults}, also demonstrated that the recognition results, for Task-$1$ ($2$-class), are better than the other two tasks, such as Task-$2$: $3$-class and Task-$3$: $4$-class, with the same number of total training images. Fig.~\ref{fig:classwiseperfm} depicts the impact of adding more classes reduces the performance metrics as such a new class increase the intra-class similarity. 
Fig.~\ref{fig:classwiseperfm} demonstrates that CVR-Net has better recognition results in binary case (Task-$1$), where it can recognize $493$-NCP samples as NCP with only $7$ samples as false negative (see in Table~\ref{tab:cm1}). Such a result is because we keep all the normal and pneumonia images as NOR class, where the intra-class similarity, between NOR and NCP classes, in Task-$1$, is very less. Whenever we break the NOR class into NOR and CPN classes in Task-$2$ ($3$-class), the recognition results are lower than the former Task-$1$ (see in Table~\ref{tab:cm1}, Table~\ref{tab:cm2}, and Fig.~\ref{fig:classwiseperfm}). Remarkably, Table~\ref{tab:cm2} depicts that the false-negative for coronavirus recognition has increased, where $15$-NCP are predicted as CPN, as well as many NOR and CPN samples are predicted as CPN and NOR, respectively. Such false-negative and false-positive results reveal that CPN class has similarities with both NOR and NCP, which CVR-Net can not recognize due to fewer samples in the training set. Further breaking of the CPN class into pneumonia bacterial (CPB) and pneumonia viral (CPV) classes, highly decreases all the metrics (see in Fig.~\ref{fig:classwiseperfm}), where, unfortunately, CVR-Net recognize the coronavirus with $40$ false negative out of $500$ samples.
Table~\ref{tab:cm3} shows that there are many false negative and false positives, which exhibit that with fewer training samples, the addition of more classes introduces the inter-class similarity and intra-class diversity. Those discussions reveal that the coronavirus recognition, from the proposed CVR-Net, is admirable, even with the less training samples, if we use less number of class as it has a minor inter-class similarity and intra-class diversity. In the future, the addition of more distinctive samples in every class, in Task-$2$ and Task-$3$, can lead CVR-Net to perform markedly even for the increased number of classes, as it has a better design to avoid the overfitting and vanishing gradient problems.   

\section{Conclusion}
\label{Conclusion}
The number of people infected by COVID-$19$ is increasing day-by-day, which can permanently damage the lungs and later provoke death. During this pandemic emergency, many countries struggle with their shortage of resources for proper recognition of the coronavirus, where such recognition, with negligible false negative, is highly essential. This article aimed to design an artificial system for automated distinguishing people with positive coronavirus. With this thing in mind, we have proposed and implemented an end-to-end deep learning-based model, called CVR-Net, to recognize the coronavirus with the very less false negative from chest radiography images without any intermediate intervention. The multi-scale-multi-encoder design of the CVR-Net ensures robustness in recognition, as the final prediction probability is the aggregation of multiple scales and encoders. In the proposed CVR-Net, different integral parts of the proposed preprocessing, such as image augmentations, class rebalancing, and transfer learning, boosted the performance of coronavirus recognition. The class rebalancing protects the model from being biased to a particular overrepresent class, as a massive number of manually annotated positive images are not publicly available yet.  The performance can further be increased by precisely segmenting the lung and adding more distinctive training samples. We also intend to deploy our trained CVR-Net to a web application for clinical utilization.


\section*{Acknowledgements}
None. No funding to declare.

\section*{Declaration of Competing Interest}
The authors have no conflict of interest to disclose.

\bibliographystyle{model2-names}

\bibliography{sample}

\end{document}